\documentclass[a4paper, 10pt, conference]{ieeeconf}
\IEEEoverridecommandlockouts                              
\usepackage[T1]{fontenc}    
\usepackage[utf8]{inputenc} 

\usepackage{graphicx} 
\graphicspath{{./figures/}}
\DeclareGraphicsExtensions{.pdf,.png,.eps,.ps}

\usepackage{amsmath,amsthm, amssymb} 

\newtheorem{theorem}{Theorem}
\newtheorem{definition}[theorem]{Definition}
\newtheorem{prop}[theorem]{Proposition}

\newcommand{\RR}{{\mathbb R}}
\newcommand{\norm}[1]{\lVert#1\rVert}

\newcommand{\otn}{{\{1...n\}}}
\newcommand{\chifree}{\chi_{\mathrm{free}}}
\newcommand{\chiobs}{\chi_{\mathrm{obs}}}
\newcommand{\chigoal}{\chi_{\mathrm{goal}}}

\title{
Optimal cooperative motion planning for vehicles at intersections
}

\author{
Jean Grégoire$^\star$\thanks{$\star$ Mines ParisTech, Centre de Robotique, Mathématiques et Systèmes, 60 Bd St Michel 75272 Paris Cedex 06, France}\hspace{10mm}
Silvère Bonnabel$^\star$\hspace{10mm}
Arnaud de La Fortelle$^{\star\dagger}$\thanks{$\dagger$ Inria Paris - Rocquencourt, IMARA team,  Domaine de Voluceau - Rocquencourt, B.P. 105 - 78153 Le Chesnay, France}
}

\begin{document}
\maketitle

\begin{abstract}
We consider the problem of cooperative intersection management. It arises in automated transportation systems for people or goods but also in multi-robots environment. Therefore many solutions have been proposed to avoid collisions. The main problem is to determine collision-free but also deadlock-free and optimal algorithms. Even with a simple definition of optimality, finding a global optimum is a problem of high complexity, especially for open systems involving a large and varying number of vehicles. This paper advocates the use of a mathematical framework based on a decomposition of the problem into  a continuous optimization part  and a scheduling problem. The paper emphasizes connections between the usual notion of vehicle priority and an abstract formulation of the scheduling problem in the coordination space. A constructive locally optimal  algorithm is proposed. More generally, this work opens up for new  computationally efficient cooperative motion planning algorithms.

\end{abstract}

\section{Introduction}

In this paper, we consider the problem of designing an intelligent transportation system meant to improve the safety at intersections, and to decrease the average time spent by vehicles going through the intersection. Due to the  achievements and promises in autonomous cars design, this subject has attracted much interest, and a variety of systems have been proposed to address this issue  \cite{Dresner2004,Hafner2011,Mehani2007,Fraichard1989,Colombo2012}. The clear improvements in the intersection flow compared to the use of traffic lights \cite{Dresner2004}, \cite{Kowshik2011}, and in terms of safety \cite{Dresner2008}, have become strong drivers for research. Moreover, conflict resolution is also an intensive field of research in railway and air transportation systems for instance \cite{Ismail1999}, \cite{Tomlin1998}.

As the task has a potentially very high complexity (essentially one tries to compactify the vehicles trajectories in space-time), the problem is often decomposed into two parts. The first part consists of determining fixed paths along which vehicles cross the intersection. The second part consists of adapting the velocity of each vehicle along the  path in order to avoid collisions --- generally resulting from human mistakes --- and to optimize as well traffic by increasing flows, i.e. by decreasing  the average time spent in the intersection. This approach was initiated in \cite{Kant1986} for multi-robot applications, and has become standard for vehicle management at intersections, e.g., \cite{Fraichard1989,Hafner2011,Leroy1999}, \cite{Akella2002}. 
 
The problem of planning velocities of robots along fixed paths has attracted much attention over the past decades, and was essentially motivated by applications for  robot manipulators and automated guided vehicles in factories. A usual approach in motion planning consists of transferring a planning problem in the physical space to a planning problem in an abstract space: the configuration space. For the problem of planning velocities along fixed paths, the configuration space is called coordination space, first introduced in \cite{ODonnell1989} for robot manipulators, and possesses a  specific cylindrical structure, as first noticed by \cite{LaValle1996}. In this paper, we propose to use the coordination space approach as a rigorous mathematical framework particularly well-suited to the problem of motion planning at intersections.  We suppose the paths of the intersection fixed and we only focus on the coordination problem.

The process is then inherently cooperative:  some vehicles need to brake to let some others go ahead until the conflict configuration is exited. The purpose of the present paper is to cast the problem in a well-defined mathematical framework, and to provide a provably collision-free solution that minimizes the average exit time over the vehicles. To do so we  revisit the notion of priority introduced in \cite{Buckley1989} and \cite{Fraichard1989}. It appears to be especially meaningful for the problem considered as we recover the usual and intuitive notion of vehicle priority at intersections. We show it is a powerful tool to classify the locally optimal trajectories that have been studied as special representatives of homotopy classes in \cite{Ghrist2006}. We introduce a priority graph and prove several properties of this graph with respect to our problem. We use it to build an algorithm for the design of collision-free and deadlock-free optimal trajectories. 

The paper has two main contributions. From an applicative viewpoint, it shows the coordination space  is a powerful mathematical tool that  is especially suited to the problem of coordinating vehicles at intersections. In particular, it opens a new research avenue for the design of efficient algorithms. From a theoretical viewpoint, our contribution consists of drawing links  between a discrete scheduling problem (priority graph) and a continuous problem formulated in the abstract coordination space. It illustrates those links in an intuitive way and relates those notions to the existence of collisions or deadlocks and to the optimality problem. New results are derived (see in particular Propositions \ref{prop1} and \ref{propOptim}) and a constructive optimal algorithm is built upon the proposed methodology.  

Section \ref{secProbForm} formulates the problem and introduces the fixed-path approach. Section \ref{secPriorRel} defines a priority relation between vehicles underlying the scheduling problem and presents some new results.  In section \ref{secBackMotionPlanProb} we present algorithms for building an optimal motion planning based on schedule decoupling.

\section{Problem formulation}
\label{secProbForm}

\subsection{Modeling assumptions}
\label{modelAsumpt}

The main assumption we make is to constrain the vehicles to follow predefined paths to go through the intersection. This assumption fits well intersections in a road network which are a highly constrained environment. Thus, every vehicle $i$ follows a particular path $\gamma_i$ and we denote $s_i \in \RR$ its curvilinear coordinate on the path. The configuration of the system of vehicles is $s=(s_i)_{i\in\otn}$ and we denote $\phi(t)$ the evolution of $s$ through time. The curvilinear coordinates are normalized and we let $\chi=[0,1]^n$ be the coordination space, where $n$ is the number of vehicles going through the intersection (possibly changing through time) and $\{\mathbf{e}_i\}_{1\le i\le n}$ the canonical basis of $\chi$. The boundedness condition on $\chi$ is rather technical but ensures the whole intersection lies in a bounded region of $\RR^n$ (somehow interactions are limited to a bounded area).

To perform motion planning optimization, we need to make hypothesis on the technical constraints of the vehicles. These include kinematic constraints (maximum velocity, maximum curve radius, etc.) and dynamic constraints (limited acceleration, adherence, jerk, etc.). In this paper, as a first step, we assume that the velocity of every vehicle on its path is limited to $v_{max}=1$ without accounting for dynamic constraints.  Moreover we assume the velocities are always positive (the cars do not go in reverse in the intersection). These assumptions are standard \cite{Fraichard1989}. 

As every vehicle occupies a non-empty geometric region, some states must be excluded to avoid collisions between vehicles. We define the obstacle region $\chiobs$ as the open set of all collision states.  A collision occurs when two vehicles occupy a same region of space, so that the obstacle region can be described as the union of $n(n-1)/2$ open cylinders $\chiobs^{ij}$ corresponding to as many collision pairs: $\chiobs = \cup_{i> j} \chiobs^{ij} $ \cite{LaValle2006}. Indeed obviously $\chiobs^{ij}=\chiobs^{ji}$ and  $\chiobs^{ii}=\emptyset$.  Figure \ref{HijFig} displays the obstacle region and a collision-free path for a two-path intersection.

Moreover, we assume that each cylinder $\chiobs^{ij}$ has an open bounded convex cross-section (in the plane generated by $\mathbf{e}_i$ and $\mathbf{e}_j$). This assumption implies cross-sections are simply connected and excludes some cases of real intersections such as a pair of opposite turn-left, but the results we present in this paper could be easily extended with a more relaxed assumption. Cross-sections are open so that the complementary set is closed and hence complete. It also ensures all cross-sections are included in the interior of $[0,1]^2$: no collision can occur for a vehicle at coordinates 0 or 1. Such assumptions are very realistic, except for the case of vehicles following each other on a same path, but our framework can be  extended to this case, as we plan to prove in future work.

\subsection{Formulation in the coordination space}\label{FormConfigSpace}
Let $\chifree=\chi\backslash \chiobs$ denote the obstacle-free space and $\mathbf{1}=(1\cdots1)$. The initial condition $x_\mathrm{init}$ belongs to $\chifree$, and the goal region is $\chigoal=\{\mathbf{1}\}\subset\chifree$. In this paper we will consider continuous paths $\phi:[0,T]\rightarrow \chi$ which are piecewise right differentiable. A collision-free path will be such that $\phi(t)\in\chi_\mathrm{free}$ for all $t$.

\begin{definition}[Feasible paths]
A feasible path for the considered  problem is a right-differentiable collision-free path such that
 \begin{itemize}
 \item $\phi(0)=x_\mathrm{init}  \quad\mathrm{and}\quad  \phi(T)\in\chi_{\mathrm{goal}}$
 \item $\phi'(t)\geq 0  \quad\mathrm{and}\quad  \norm{\phi'(t)}_\infty\leq 1\qquad \forall t\in [0,T]$.
 \end{itemize}
\end{definition}

In order to define an optimality criterion for the motion planning problem, we need to define a cost function $c(\phi)$ enabling to compare feasible paths. As explained in the introduction, the purpose is to minimize the average exit time over the vehicles. As a consequence, we define the cost function $c$ as:
$$
c(\phi)=\frac{1}{n}\sum_{i=1}^n T_i = \frac{1}{n}\sum_{i=1}^n \phi_i^{-1}(1)
$$
with $\phi_i^{-1}(s_i)$ denoting the first date at which $\phi_i$ reaches $s_i$ and $T_i=\phi_i^{-1}(1)$ being the exit time for the vehicle $i$.

\begin{figure}[ht]
\centering
\includegraphics[width=1\linewidth]{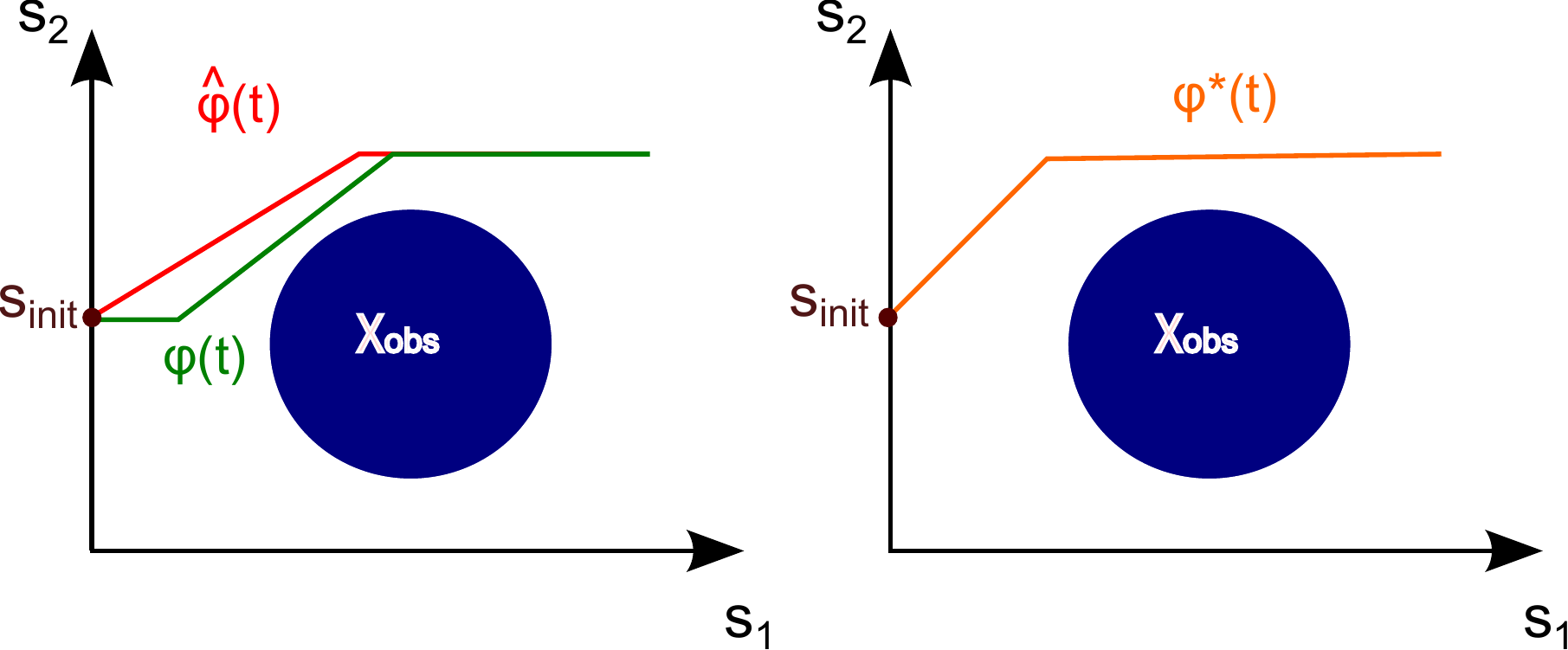}\hfill
\caption{The left drawing depicts two feasible trajectories in a two-dimensional coordination space. With the trajectory $\hat\phi$, the vehicle $2$ leaves the intersection earlier than with $\phi$, while exit time is not changed for the vehicle $1$. Hence, $c(\hat\phi) < c(\phi)$. $\phi^*$ is an optimal trajectory since every vehicle is always at maximum speed: $\phi^* = \arg\min c(\phi)$.}
\label{costExamples2D}
\end{figure}

The optimality problem consists of finding a feasible path $\phi^*$ that minimizes the cost, i.e. for any feasible path $\phi$, we have $c(\phi^*)\leq c(\phi)$. Figure \ref{costExamples2D} depicts three feasible trajectories with different costs in a two-dimensional coordination space for a given initial condition.

Note that the cost can be easily bounded from above and from below. Suppose for simplicity's sake that  $x_\mathrm{init}=0$ and $\chi_{\mathrm{goal}}=\{\mathbf{1}\}$. If $\chiobs=\emptyset$ the optimal path is given by $\phi'\equiv\mathbf{1}$. The corresponding cost $\underline c(\phi)=\frac{1}{n}\sum_{i=1}^n 1=1$ is thus a lower bound on the optimal time in the general case. On the other hand, a suboptimal trajectory is easily found moving the vehicles one after the other, i.e. moving along the edges of the hypercube. This trajectory is collision-free, because by assumption a vehicle cannot collide another one if its coordinate is $0$ or $1$. It yields  $\bar c(\phi)=\frac{1}{n}\sum_{i=1}^n i=\frac{n+1}{2}$.

Note that vehicle $i$ is free after $T_i$ when its coordinate is 1: there is no further interaction with any other vehicle. We hereafter take $\phi_i(t)=1$ for $t\ge T_i$ in order to stay in $\chi=[0,1]^n$ but any further trajectory $\phi_i(t)\ge 1$ would be suitable had we taken $\chi=\RR^n$.

\subsection{Geometry of the coordination space}

In this section, we propose several geometric concepts in the coordination diagram.

\subsubsection{Completing the obstacle region}

\begin{figure}[ht]
\centering
\includegraphics[width=1\linewidth]{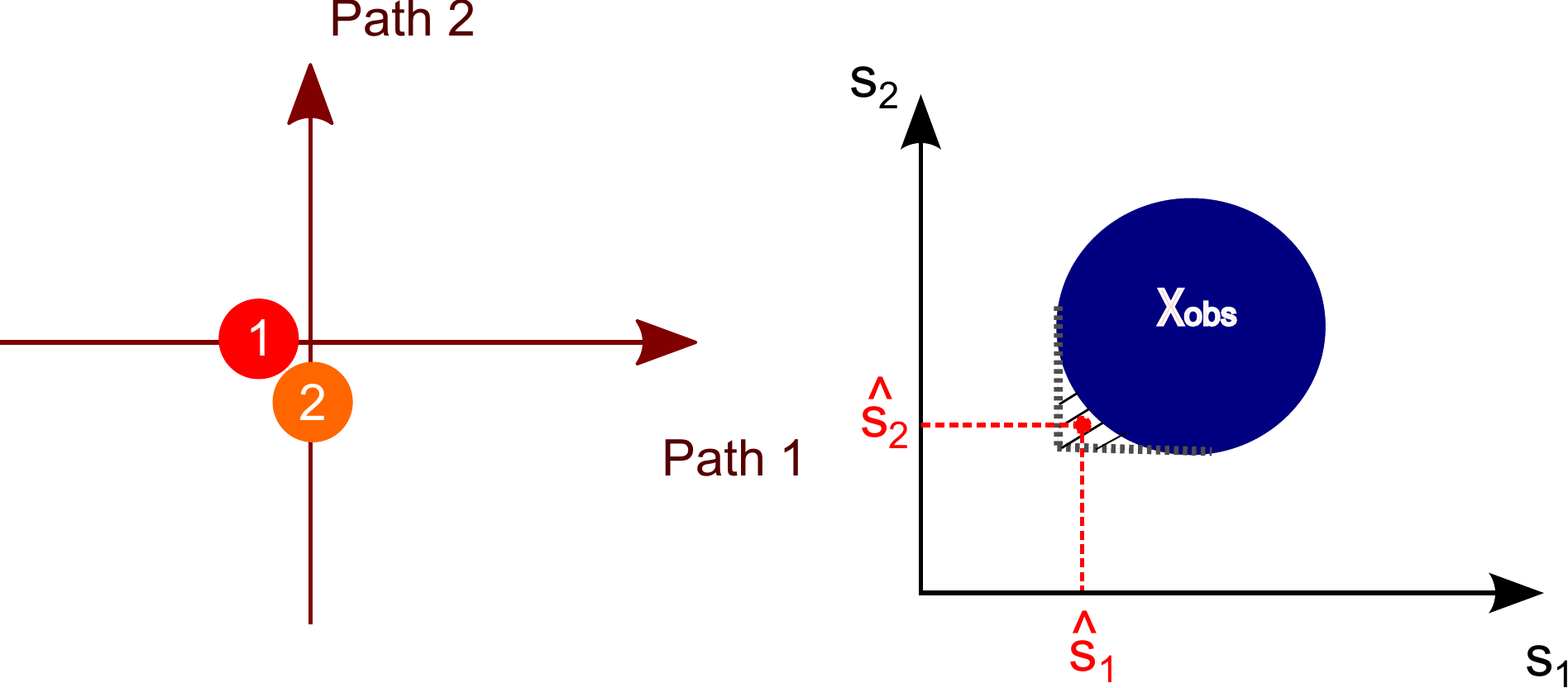}\hfill
\caption{Two representations of a dead-lock situation involving two circle-shaped vehicles. The left drawing depicts the real intersection with its two orthogonal paths and the two circle-shaped vehicles. Both of them have moved forward too much, so that now they cannot exit the intersection without turning around. The right drawing represents what happens in the two-dimensional coordination space and highlights the South-West region that must be added to $\chiobs$ to avoid the one-to-one dead-lock.}
\label{oneToOneDeadLock}
\end{figure}

For each collision pair $(i,j)$ we define the south of $\chiobs$ along the plane $\mathbf{e}_i \times \mathbf{e}_j$ as:
\begin{eqnarray}
\label{south.def}
\mathrm{S}_{ij}(\chiobs) &=& \chiobs^{ij}-\RR_+ \mathbf{e}_j\\
\nonumber
  &=& \left\{s-\lambda \mathbf{e}_j:  \forall s\in \chiobs^{ij},\ \forall \lambda\in\RR_+ \right\}
\end{eqnarray}
We say the south of $\chiobs^{ij}$ along the plane $\mathbf{e}_i \times \mathbf{e}_j$ is the west of $\chiobs^{ij}$ along the plane $\mathbf{e}_j \times \mathbf{e}_i$ and denote $\mathrm{W}_{ij}(\chiobs)=\mathrm{S}_{ji}(\chiobs)$. These sets are clearly open. They  play the role of  gates: a feasible trajectory must go through one or the other and they are exclusive. As a result, the intersection of the south and the west is forbidden for feasible trajectories, even if there is no collision. As the velocities must be positive, being in the intersection at some time would necessarily lead to a future collision unless the cars stop moving (see Figure \ref{oneToOneDeadLock}). This situation is referred to as a deadlock. To avoid such situations, an extended obstacle region is defined, so that no basic one-to-one dead-lock situation can occur \cite{ODonnell1989}. 

\begin{definition}[SW--Completion]
The South-West completion of $\chiobs$ along the plane $\mathbf{e}_i \times \mathbf{e}_j$ is given by
$$
\mathrm{SW}_{ij}(\chiobs)= \mathrm{S}_{ij}(\chiobs)\cap \mathrm{W}_{ij}(\chiobs)
$$
where $\mathrm{S}_{ij}(\chiobs)$ is defined in~(\ref{south.def}).
The completed obstacle region $\overline{\chiobs}$ is defined by
$$
\overline{\chiobs} = \bigcup_{i\neq j} \mathrm{SW}_{ij}(\chiobs)
$$
We call $\overline{\chifree} = \chi \backslash \overline{\chiobs}$ the obstacle-free completed region.
\end{definition}

The South-West closure is applied on each cylinder $\chiobs^{ij}$ as depicted on Figure \ref{oneToOneDeadLock}. An important fact --- not explicitly mentioned in the literature yet --- is that  this method cannot be readily extended to avoid dead-locks in multi-dimensional planning problems while ensuring optimality. For this reason, the next sections rather present tools based on gates' positions to avoid dead-locks caused by multiple vehicles' interaction.

\subsubsection{Gates in the coordination space}\label{gates}

Let define the region $H_{i \succ j}$ as follows:
$$
H_{i\succ j} = \mathrm{S}_{ij}(\chiobs) \setminus \mathrm{SW}_{ij}(\chiobs)
$$

\begin{figure}[ht]
\centering
\includegraphics[width=1\linewidth]{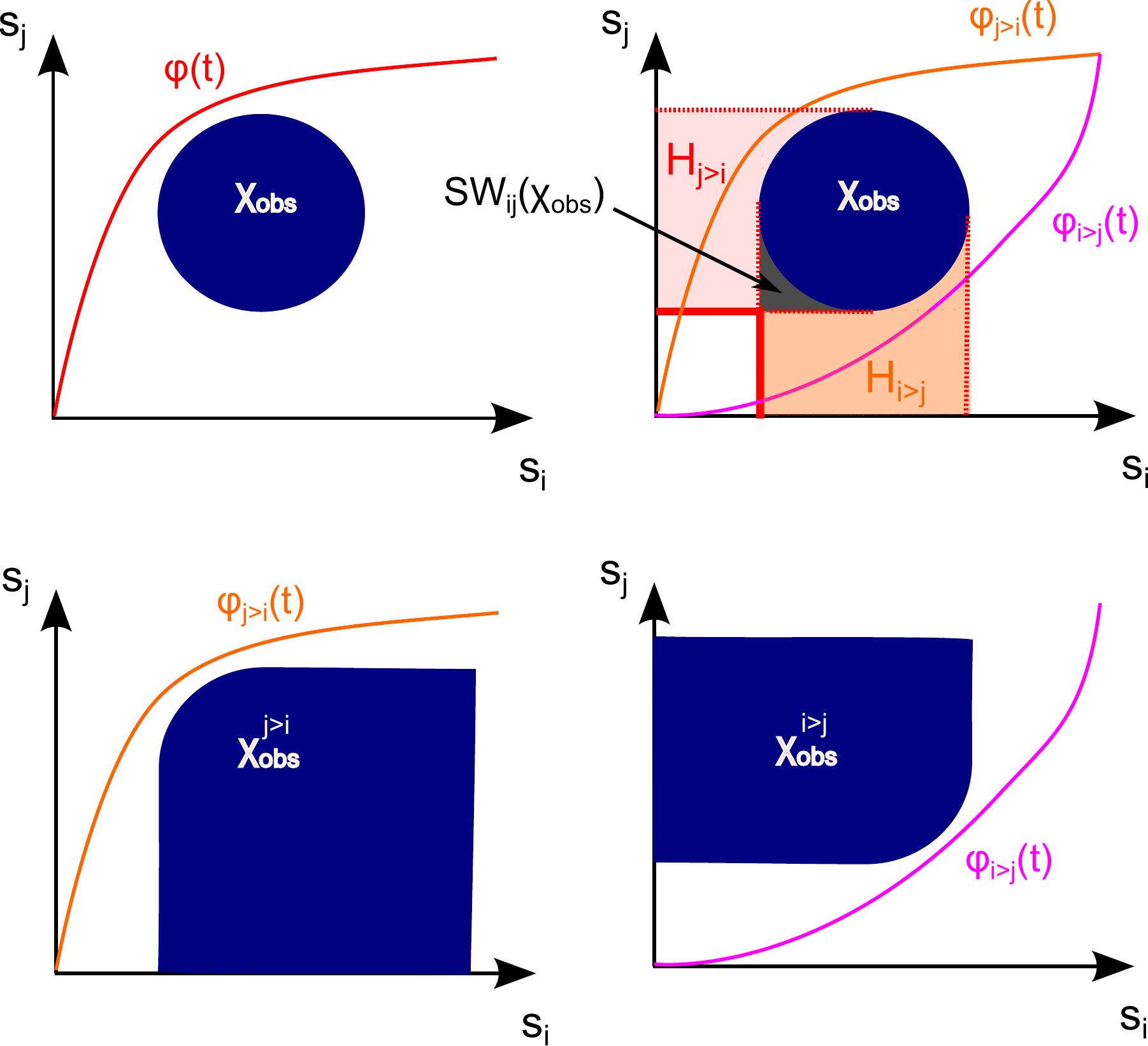}\hfill
\caption{The top-left drawing represents in the plane $(s_i,s_j)$ the obstacle region $\chiobs$ and a feasible trajectory $\phi$. The top-right drawing visualizes the two gates $H_{i\succ j}$ and $H_{j\succ i}$. For each gate, a feasible trajectory passing through the gate is given. The bottom drawings depict $\chiobs^{i\succ j}$ and $\chiobs^{j\succ i}$.}
\label{HijFig}
\end{figure}

The top drawings of Figure \ref{HijFig} display $H_{i \succ j}$ and $H_{j \succ i}$. Over the convexity hypothesis of the cylinders of $\chiobs$, we can assert that any feasible trajectory will necessarily cross $H_{i \succ j}$ or $H_{j \succ i}$ exclusively. This is why these two boxes can be viewed as gates in the coordination space. In Figure \ref{HijFig}, the path $\phi_{i\succ j}$ intersects the gate $H_{i \succ j}$: the vehicle $j$ slows down to let the vehicle $i$ go ahead. Each gate favours a vehicle over another i.e. a priority is assigned, and passing through a gate possibly prevents from passing through many other gates because of the positive speed assumption.

We propose to define the obstacle region $\chiobs^{i \succ j}$ as a set that is incompatible with the crossing of gate  $H_{i \succ j}$, as illustrated on Figure \ref{HijFig}: $$\chiobs^{i\succ j} = \chiobs^{ij} - \RR_+ \mathbf{e}_i + \RR_+ \mathbf{e}_j$$

\section{Priority relations and ordering}
\label{secPriorRel}

The geometry of the coordination space leads us to define a natural binary relation corresponding to priority relations between vehicles: a  very familiar and intuitive notion in real life. Indeed, we say the vehicle $i$ has priority over the vehicle $j$ in the intersection if the associated path goes through the gate $H_{i \succ j}$ in the coordination space.
\begin{definition}[Priority relation]
A feasible path $\phi$ induces a binary relation $\succ$ on the set $\otn$ as follows. For $i\neq j$, $i \succ j$ if $\phi$ intersects $H_{i \succ j}$.
\end{definition}

The following proposition proves a result that is not totally obvious from an intuitive viewpoint.
\begin{prop}
The binary relation $\succ$ induced by a feasible trajectory $\phi$ does not necessarily defines an order.
\end{prop}

\begin{figure}[ht]
\centering
\includegraphics[width=.4\linewidth]{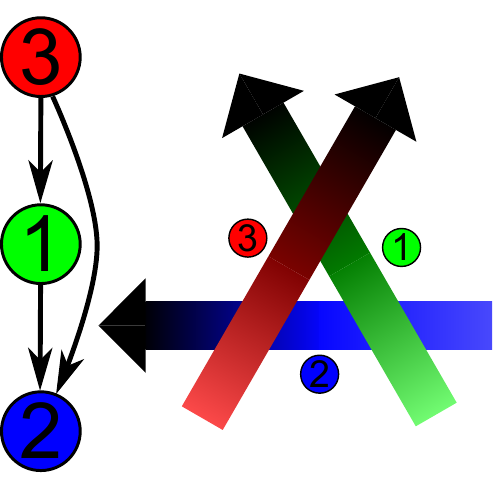}\hfill
\includegraphics[width=.4\linewidth]{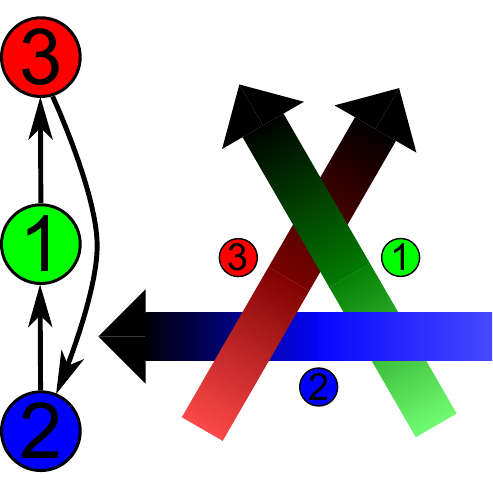}
\caption{Two representations of priority relations. In each drawing relation is represented in two ways: as a complete oriented graph, where orientation yields the priority; and as trajectories over time, foreground being first, background later. The left drawing represents a relation that  is an order (even a total order). The right drawing shows a relation that is \emph{not} an order.}
\label{priorite-1}
\end{figure}

Figure \ref{priorite-1} depicts a 3-paths intersection with 3 intersection points and as many vehicles going through the intersection. In the right drawing, the red vehicle goes ahead the blue vehicle and the green vehicle goes ahead the red vehicle, while the blue vehicle goes ahead the green vehicle. It yields in a non transitive binary relation: the set $\otn$ is not partially ordered.

\subsection{Priority graph}

In this subsection, we propose to define the priority relations in terms of a graph. As for any collision pair $i,j$ we have either $i \succ j$ or $j \succ i$, the priority relation can be defined by an oriented graph $G$ whose vertices are $\otn$.

\begin{definition}
Given a feasible trajectory $\phi$, we call the priority graph the oriented graph $G$ whose vertices are $\otn$ and such that $i \xrightarrow{G} j$ if and only if $i \succ j$.
\end{definition}

Note the graph is complete in the sense that for all $i\neq j$ there is one, and only one, arc linking $i$ and $j$. There is no arc linking $i$ to itself. Generally speaking, there are potentially $2^{\frac{n(n-1)}{2}}$ possible priority graphs. Actually, one can easily understand that some priorities are not acceptable. The next section deals with this issue and characterizes what we call feasible priorities.

\subsection{Feasible priorities}

\begin{definition}[Feasible Priority]
A priority graph $G$ is feasible if there exists a feasible trajectory $\phi$ whose priority graph is $G$.
\end{definition}

\begin{figure}[ht]
\centering
\includegraphics[width=1\linewidth]{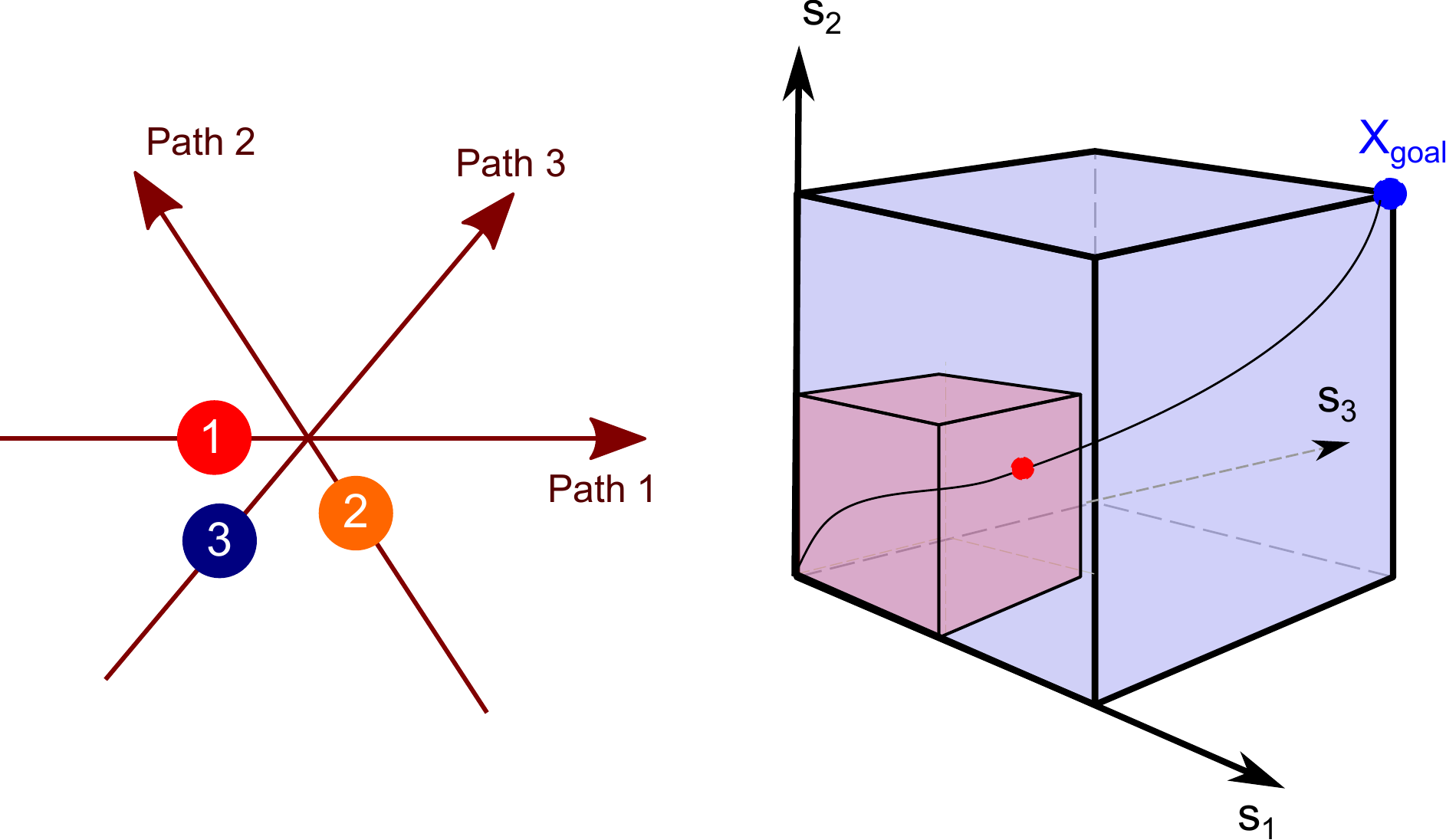}\hfill
\caption{The left drawing represents a 3-path intersection with a common intersection point and 3 vehicles going through the intersection. The right drawing depicts what happens in the coordination space when cyclic priorities are set. Any feasible trajectory respecting cyclic priorities should stay in the red box $[0,\frac{1}{2}]^3$, but reaching the goal requires to cross this box.}
\label{3PathExample}
\end{figure}

In the intersection depicted on the left drawing of Figure \ref{3PathExample}, we can assert that some priorities are not feasible. Indeed, if the the vehicle 2 goes ahead the vehicle 1 and the vehicle 2 goes ahead the vehicle 3, the vehicle 1 must go ahead the vehicle 3, because the three paths have a common intersection point. Hence, for this particular situation, the priority graph cannot be cyclic and the priority relation is an order relation.

The proposition below provides a necessary condition for priorities feasibility based on the relative position of the fixed-priority collision cylinders $\chiobs^{i \succ j}$. The proof
is omitted due to space limitations.

\begin{prop}[Characterization of feasible priority graphs]\label{prop1}
A priority graph $G$ is feasible if and only if for any cycle $\mathcal{C}$ in $G$, $$\bigcap_{i \xrightarrow{\mathcal{C}} j}  \chiobs^{i \succ j} = \emptyset$$
\end{prop}

As $\chiobs^{ij} \subset \chiobs^{i \succ j}$ it readily implies:

\begin{prop}[Necessary condition for priorities feasibility]\label{prop2}
Suppose $G$ is a feasible graph. Then, for any cycle $\mathcal{C}$ in $G$, $$\bigcap_{i \xrightarrow{\mathcal{C}} j}  \chiobs^{ij} = \emptyset$$
\end{prop}

In the intersection of Figure \ref{3PathExample}, the state $(s_i=\frac{1}{2})_{i=1,2,3}$ corresponding to a common intersection point is in each of the three collision cylinders. Hence, the necessary condition applies and it can be asserted that no cycle between the 3-vehicle priority graph can exist. The underlying reason in the coordination space is that if there is a cycle in the priorities, then the trajectory must stay in the box $[0,\frac{1}{2}]^3$ at any time to respect these priorities, i.e. cannot reach $\chigoal$.

The characterization of feasible priorities shows that the feasibility of a priority graph relies on the relative position of the collision cylinders in the coordination space.

\section{The motion planning problem}
\label{secBackMotionPlanProb}

The last section shows that the set of feasible trajectories can be divided into as many disjoint subsets as there are feasible priority graphs. In this section, we shall present an algorithm that builds an optimal trajectory for a fixed priority graph. We will then discuss the computational issues for determining the globally optimal trajectory.

\subsection{Optimal trajectories with fixed priority graph}

\label{algoOptPrioFix}

The main idea  is that when priorities are fixed, if at any time every vehicle runs as much distance as possible in a time step, then we can assert that the trajectory is optimal (i.e.  minimizes the average time spent in the intersection) for the fixed priorities. It means the optimal solution can be defined reactively. 

Let  $H_\mathrm{G}$ denote the union of gates corresponding to the priorities fixed by $G$. After setting $\phi(0)=x_{\mathrm{init}}$, the algorithm proceeds as follows. If $G$ is not a feasible priority graph, the algorithm stops, else, until $\chigoal$ is reached, at each date $t$, the trajectory is defined via $\phi'(t)$ which is calculated as a function of $\phi(t)$ by the following algorithm, that essentially consists of projecting the vector $\mathbf{1}$ consecutively:
\begin{itemize}
\item $\phi'(t)\leftarrow\mathbf{1}$
\item \textit{Stay in $\chi$}: For $i$ such that $\phi_i(t)=1$, do $\phi_i'(t)\leftarrow 0$
\item \textit{Avoid Forbidden Gates}: While there exists $i \xrightarrow{G} j$ such that $\phi(t)$ belongs to the boundary of the gate $H_{j \succ i}$, do: \\$\phi'(t)\leftarrow$ orthogonal projection of $\phi'(t)$ on the boundary of $H_{j \succ i}$
\item \textit{Avoid Obstacle Region}: While there exists $i \xrightarrow{G} j$ such that $\phi(t)$ belongs to the boundary of $\overline{\chiobs}^{i j}$, if $\mathbf{1}$ points towards $\overline{\chiobs}^{i j}$, do: \\ $\phi'(t)\leftarrow$ orthogonal projection of $\phi'(t)$ on any tangential hyperplane to $\overline{\chiobs}^{i j}$ at $\phi(t)$
\end{itemize}

\begin{prop}[Optimal motion planning for fixed priorities]
\label{propOptim}
Given a feasible priority graph $G$, the trajectory $\phi^*(t)$ generated by the algorithm described above is an optimal trajectory for the given priorities.
\end{prop}

\begin{figure}[ht]
\centering
\includegraphics[width=1\linewidth]{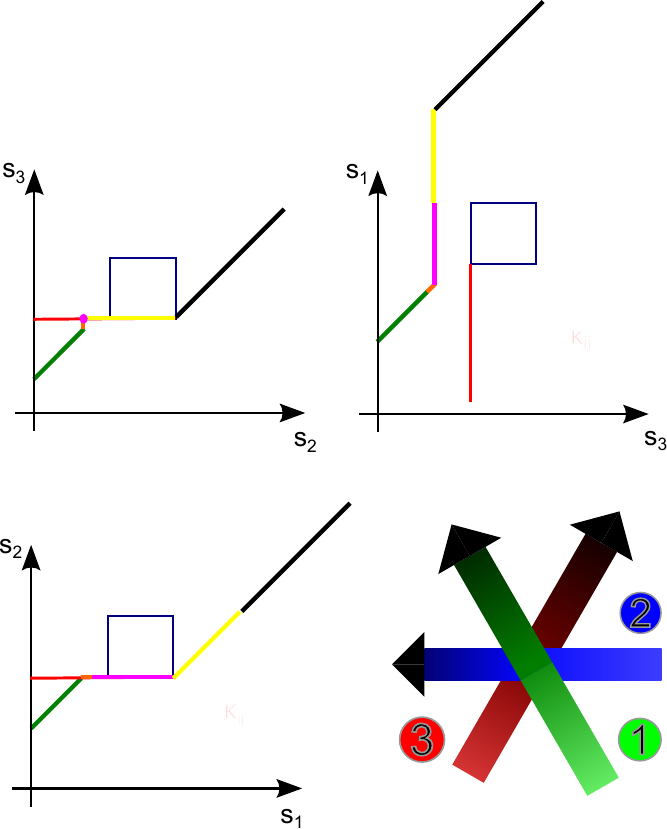}\hfill
\caption{The bottom-left drawing represents a 3-path intersection where priorities are fixed to $1 \succ 2$,  $2 \succ 3$ and $1 \succ 3$. The three other drawings visualize in each plane $(s_i,s_j)$  the obstacle region (in blue), the forbidden gates (red), and the optimal trajectory generated by the algorithm. The color of the trajectory changes when a collision pair leaves or reaches a gate or the boundary of the obstacle region.}
\label{optimalExample}
\end{figure}

It can be noted that the vehicles may brake for two reasons: avoiding the obstacle region or avoiding a gate that is not compatible with the priorities defined by $G$. Figure \ref{optimalExample} displays a trajectory generated by the algorithm for a 3-paths intersection. For simplicity's sake, for each collision pair, the one-to-one obstacle region is approximated to a square.

\begin{proof} Suppose there exists a trajectory $\psi$ respecting the priorities $G$ and the initial condition $\psi(0)=\phi^*(0)$ such that $\psi_{i_0}$ overtakes $\phi^*_{i_0}$ for the first time at the date $t_0>0$. Consider the trajectory defined by $\phi_i(t)=\max(\psi_i,\phi^*_i)(t)$. The trajectory $\phi$ admits $G$ as priority graph, it is collision-free (because $\psi$ and $\phi^*$ respect the same priorities, see Figure \ref{HijFig}) and it respects the maximum velocity constraint. Moreover, $\phi_i(t_0)=\psi_i(t_0)$ and for small enough $\epsilon>0$, $\phi_{i_0}(t_0+\epsilon) \geq \phi^*_{i_0}(t_0)$. By construction of $\phi^*$, for small enough $\epsilon, \phi_i(t_0+\epsilon) \leq \phi^*_i(t_0)$, so that the last two conditions require $\phi_{i_0}(t_0+\epsilon)=\phi^*_{i_0}(t_0+\epsilon)=\psi_{i_0}(t_0+\epsilon)$ to respect both inequalities. As a consequence, for all $i\in\otn$, the trajectory $\phi^*_i$ is never strictly overtaken by a vehicle in another feasible trajectory $\psi_i$: $\phi^*$ is an optimal trajectory.
\end{proof}

The optimal solution constructed by the algorithm is known as the "left-greedy" solution  \cite{Ghrist2006}. In this latter paper the authors address a more general problem of optimization in coordination space, and regroup the trajectories in equivalence classes based on homotopy (two trajectories in the same class can be deduced from each other by a continuous deformation). In our formalism, each admissible graph defines an homotopy class. In our particular application, we recover their optimality result with simple means (as opposed to their proof that is based on concepts of advanced differential geometry, namely Gromov's link condition for CAT(0) spaces). 

\subsection{Optimal trajectories}

As the algorithm of section \ref{algoOptPrioFix} generates an optimal trajectory for a given priority graph, the motion planning optimization boils down to a combinatorial problem. By computing and comparing the costs of the optimal trajectories among the feasible priority graphs, we can conclude on the optimal planning solution by choosing the least costly trajectory. This method is suitable when there are few vehicles crossing the intersection, but its high algorithmic complexity (due to the comninatorics of priorities) can become a prohibitive numerical cost when the number of vehicle grows. This motivates the following heuristic algorithm. 

\subsection{A heuristic algorithm with unconstrained graph}
The last section proves that once the priority graph is fixed, the optimum is reached  following algorithms belonging to the  "bug"  family,
emanating  from the work  of \cite{Lumelsky1990}. Indeed, the optimal paths go at maximum speed until they reach the frontier of a forbidden region. Then, they follow the boundary as long as necessary. In the absence of fixed priority graph, we propose to have the same reactive approach, moving the vehicles as fast as  possible. A graph is thus iteratively built, and  each time a gate is reached, the retained priority must be compatible with the graph already built.

After setting $\phi(0)=x_\mathrm{init}$ and initializing $G$ to an empty graph, the algorithm proceeds as follows until $\chigoal$ is reached:
\begin{itemize}
\item $\phi'(t)\leftarrow\mathbf{1}$
\item \textit{Stay in $\chi$}: For $i$ such that $\phi_i(t)=1$, do $\phi_i'(t)\leftarrow 0$
\item \textit{Update $G$}: While there exists a collision pair $i,j$ whose relative priorities are not defined in $G$ and such that $\phi(t)$ belongs to the boundary of $H_{i \succ j}$, do $i \xrightarrow{G} j$ if it yields a feasible priority graph, else do $j \xrightarrow{G} i$
\item \textit{Avoid Forbidden Gates}: While there exists $i \xrightarrow{G} j$ such that $\phi(t)$ belongs to the boundary of the gate $H_{j \succ i}$, do: \\$\phi'(t)\leftarrow$ orthogonal projection of $\phi'(t)$ on the boundary of $H_{j \succ i}$
\item \textit{Avoid Obstacle Region}: While there exists $i \xrightarrow{G} j$ such that $\phi(t)$ belongs to the boundary of $\overline{\chiobs}^{i j}$, If $\mathbf{1}$ points towards $\chiobs^{ij}$, do: \\ $\phi'(t)\leftarrow$ orthogonal projection of $\phi'(t)$ on any tangential hyperplane to $\overline{\chiobs}^{i j}$ at $\phi(t)$
\end{itemize}
The trajectory is defined recursively, the algorithmic complexity remains relatively low, and yields a generated priority graph $G$ \emph{a posteriori}. Moreover, we have guarantees the built trajectory is optimal for this priority graph.

\section{Conclusion}

The mathematical framework we propose to address the cooperative motion planning problem at intersections is based on path-velocity decomposition. Hence, the problem boils down to searching an optimal path in a high-dimensional coordination space. We define a priority relation between the vehicles that provides a tool to classify the feasible trajectories, and propose an algorithm to construct
an optimal trajectory for given priorities. 

As the search for a global optimum is inherently related to the determination of all optima given a graph, it is of high algorithmic complexity. As a result, a good heuristics for fixing priorities is at the core of a simple quasi-optimal solution, as suggested by the intuition.

In the future, we plan to find such heuristics, and to take into account more realistic assumptions: varying number of vehicles, more general coordination space, and above all dynamic constraints. The next step will be the integration of the selected algorithms in a real intersection crossed by numerous vehicles such as cooperative cybercars developed at INRIA.

\begin{figure}
\includegraphics[width=\linewidth]{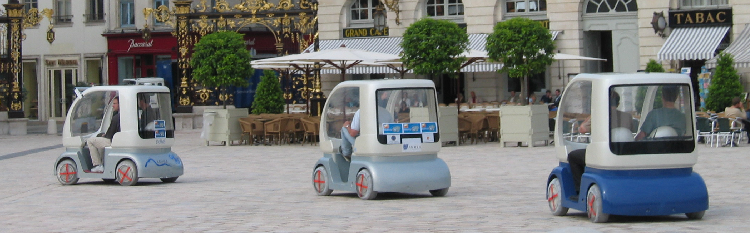}
\caption{Cooperative cybercars in the French city of Nancy during the MobiVIP project: INRIA develops fully automated transportation systems that need cooperative motion planning (here is a platooning example).}
\label{cycabs}
\end{figure}

\bibliographystyle{plain}
\bibliography{biblio}

\end{document}